\documentclass[aps,prl,twocolumn,showpacs,superscriptaddress, amsmath,amssymb,tightenlines]{revtex4} 
\newcommand{\ie}{i.e. \ } 
\newcommand{\eg}{e.g. \ }

\usepackage{graphicx} 

\usepackage[a4paper,left=1.5cm,right=1cm,top=3cm,bottom=2cm]{geometry}
 
\begin{document} 
 
\title{Bismuth qubits in silicon: the role of EPR ``cancellation resonances''} 
\author{M. H. Mohammady}
\affiliation{Department of Physics and Astronomy, 
University College London, Gower Street, London WC1E 6BT, United Kingdom} 
\author{G. W. Morley}
\affiliation{Department of Physics and Astronomy, 
University College London, Gower Street, London WC1E 6BT, United Kingdom}
\affiliation{London Centre for Nanotechnology,
University College London, Gordon Street, London WC1H 0AH, United Kingdom}
\author{T. S. Monteiro}
\affiliation{Department of Physics and Astronomy, 
University College London, Gower Street, London WC1E 6BT, United Kingdom}
\date{\today} 
\begin{abstract} 
 
We investigate theoretically and experimentally the electron paramagnetic resonance (EPR)
spectra of bismuth doped silicon (Si:Bi) at intermediate magnetic 
fields, $B \approx 0.05 -0.6$ T. We identify a previously unexplored EPR regime of  
``cancellation-resonances''- where the non-isotropic part of $ AS_zI_z$, the Ising part of
the hyperfine coupling, is resonant with the external field-induced splitting. 
 We show this regime has interesting and 
experimentally accessible consequences for spectroscopy and quantum information applications.
These include reduction of decoherence,  
fast manipulation of the coupled nuclear-electron qubit system
and line narrowing in the multi-qubit case.
 We test our theoretical analysis by comparing with 
 experimental X-band (9.7 GHz)  EPR spectra obtained in the intermediate field regime.     
 
\end{abstract} 
\pacs{03.67.Lx,03.67.-a,76.30-v,76.90.+d,} 
\maketitle 
  
Following Kane's suggestion \cite{Kane} for using phosphorus doped silicon as a source of  
qubits for quantum computing, there has been intense interest in such systems \cite{SiP}. 
The phosphorus system ($^{31}$P) is appealing in its simplicity: it
 represents a simple electron-spin qubit $S=\frac{1}{2}$ coupled to a nuclear-spin qubit
 $I=\frac{1}{2}$ via an  
isotropic hyperfine interaction $A {\bf I}.{\bf S}$ of moderate strength  
($\frac{A}{2\pi}=117.5$ MHz). 
However, recent developments \cite{LCN,Thewalt,Morton}
point to Si:Bi (bismuth doped silicon) as a very promising new alternative.
Two recent studies measured spin-dephasing times of over 1 ms  at $10$ K which 
is longer than  comparable (non-isotopically purified) materials, including
 Si:P \cite{LCN, Morton}. Another group implemented a scheme for 
rapid (on a timescale of $\sim 100 \mu$s) and efficient (of order $90\%$) 
 hyperpolarization of Si:Bi into a single spin-state \cite{Thewalt}. 
 
Bismuth has an atypically large hyperfine 
constant $\frac{A}{2\pi}=1.4754$ GHz and nuclear spin $I=\frac{9}{2}$. This makes its EPR spectra somewhat 
more complex than for phosphorus and there is strong mixing of the eigenstates for  
external field $B \lesssim 0.6$ T. Mixing of Si:P states was studied
 experimentally in \cite{Itoh}, by means
 of electrically detected  magnetic resonance (EDMR), but at much lower fields $B \lesssim 0.02$ T.
 Residual mixing in Si:Bi for $B=2-6$ T, where the eigenstates are  
$\gtrsim 99.9\%$ pure uncoupled eigenstates of both ${\hat I}_z$ and ${\hat S}_z$, was 
also proposed as important for the  hyperpolarization mechanism of illuminated Si:Bi   
 \cite{Thewalt}.  In \cite{Morton}
it was found that even a $\sim 30\%$ reduction in the effective paramagnetic 
ratio $\frac{df}{dB}$ (where $f$ is the transition frequency) lead to a detectable reduction
in decoherence rates. 

Below we present an analysis of EPR spectra for Si:Bi and test the results with
experimental spectra. We identify a series of regimes for which $\frac{df}{dB}=0$,
 explaining them in a unified manner as a series of  EPR ``cancellation
resonances''; some are associated with avoided level-crossings
while others, such as a maximum shown in 
ENDOR \cite{ESEEM} spectra at $B \approx 0.37$ T in  \cite{Morton} is of a quite different origin.
These cancellation resonances represent, to the best of our knowledge,
 an unexplored regime in EPR spectroscopy, arising in systems with exceptionally high $A$ and $I$.
They are somewhat reminiscent of the so-called ``exact cancellation'' regime, widely
used in ESEEM spectroscopy \cite{ESEEM,Schweiger}, but differ in essential ways:
 they affect both electronic and
nuclear frequencies rather than only nuclear frequencies; they concern only
the non-isotropic component of the interaction (and are thus not ``exact'';
indeed the $B \approx 0.37$ T point is not even a full cancellation). They have important
implications for the use of Si:Bi as a coupled electron-nuclear qubit pair: we show
all potential spin operations may be carried
out with fast EPR pulses (on nanosecond timescales) where in contrast,  most operations
for Si:P require slower NMR pulses (on microsecond timescales). A striking spectral
signature is reduced sensitivity to certain types of ensemble averaging, giving an
analog to the ultra-narrow lines well-known in ``exact cancellation'', as well
as the reduction of decoherence. Further details are found in \cite{long}.

 \begin{figure}[tb] 
\includegraphics[width=3in]{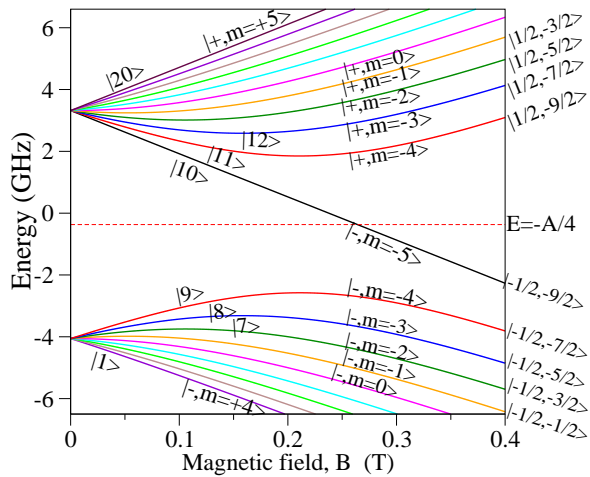} 
\caption{The 20 spin energy-levels of Si:Bi may be labelled in alternative ways:
{\bf (1)} in order of increasing energy $|1\rangle,|2\rangle ...|20\rangle$
{\bf (2)} using the adiabatic basis $|\pm,m\rangle$ of doublets; levels belonging to the same
   doublet are shown in the same color (online).
 {\bf (3)} by their asymptotic, high-field form $|m_s,m_I\rangle$.
States $|10\rangle$ and $|20\rangle$ are not mixed. State $|10\rangle$ is of especial
significance since it (rather than the ground state) is a favourable state to 
initialise the system in (experimental hyperpolarization studies \cite{Thewalt}
 concentrate the system in this state). Thus, in our coupled 2-qubit scheme, state
$|10\rangle$ corresponds to our $(0)_e(0)_n$ state; in the same scheme,
 states $|9\rangle \equiv (0)_e(1)_n$ and  $|11\rangle \equiv (1)_e(0)_n$
are related to $|10\rangle$ by a single qubit flip (of either the electron or nuclear
qubits respectively) while for $|12\rangle \equiv (1)_e(1)_n $ both qubits are flipped.}
\label{Fig1} 
\end{figure} 

We model the Si:Bi spin system approximately by a Hamiltonian  
including an isotropic hyperfine coupling term: 
\begin{equation} 
{\hat H}= \omega_0 {\hat S}_z - \omega_0 \delta {\hat I}_z  + A {\bf{\hat S} . {\hat I}} 
\label{HAM} 
\end{equation}    
where $\omega_0$ represents the frequency of the external field and  
$\delta=\omega_I/\omega_0 =2.488\times 10^{-4}$ represents 
the ratio of the nuclear to electronic Zeeman frequencies. For $I=\frac{9}{2}$, $S=\frac{1}{2}$ there  
are 20 eigenstates which can be superpositions of high-field eigenstates $|m_s,m_I\rangle$;
but since $[{\hat H},{\hat S}_z+{\hat I}_z]=0$, the $|m_s,m_I\rangle$ basis   
 is at most mixed into a doublet  $|m_s=\pm\frac{1}{2} ,m_I=m\mp\frac{1}{2}\rangle$
with constant $m=m_s+m_I$.  One can thus write the Hamiltonian for each $m$ sub-doublet
 as a $2$ dimensional matrix ${H}_m$ 
(where $H_m = \frac{A}{2}{\tilde h}_m$):
\begin{equation} 
{\tilde h}_m=[m+{\tilde \omega_0}(1+ \delta)]{\hat \sigma}_z  + (25-m^2)^{1/2}{\hat \sigma}_x 
              -(\frac{1}{2}+ 2m\delta{\tilde \omega_0}){\bf I}
\label{Hm} 
\end{equation}      
and where ${\tilde \omega_0}= \frac{\omega_0}{A}$ is the rescaled field, 
  ${\hat \sigma}_z,{\hat \sigma}_x$ represent Pauli matrices 
in the two-state basis $|m_s,m_I\rangle= |\pm \frac{1}{2}, m\mp \frac{1}{2}\rangle$
and ${\bf I}$ is the identity operator. It becomes clear that whenever  
$m =- {\tilde \omega_0}(1+ \delta)$, the quantum states become 
eigenstates of ${\hat \sigma}_x$. Thus at $m \approx -{\tilde \omega_0}$, the 
eigenstates, $|\pm, m\rangle$, assume Bell-like form: 
 ( $|\pm, m\rangle=\frac{1}{\sqrt{2}}\left[|-\frac{1}{2}\rangle|m+\frac{1}{2}\rangle \pm
 |+\frac{1}{2}\rangle |m-\frac{1}{2}\rangle \right]$).  In contrast,  ``exact cancellation''
results in a simple superposition of the nuclear states, 
which also allows other types of manipulations \cite{Mitrikas}. Since $\omega_0 \geq 0$, 
only states with $-5 \leq m \leq 0$  can yield resonances where the field-splitting 
term $\frac{1}{2}\omega_0 {\hat \sigma}_z$ is eliminated.
 In this case, they occur at  
 $\omega_0 \simeq 0,A,...4A,5A$, corresponding to applied field $B=0,0.053,...0.21,0.26$ Tesla;
below we show that all $\frac{df}{dB}=0$ points which are minima occur midway between these resonances.
 But the $\frac{df}{dB}=0$ maximum at the $\omega_0=7A \equiv 0.37$ T resonance, 
and seen in experiments \cite{Morton} is shown to be of a different type.

\begin{figure}[htb] 
\includegraphics[width=2.7in]{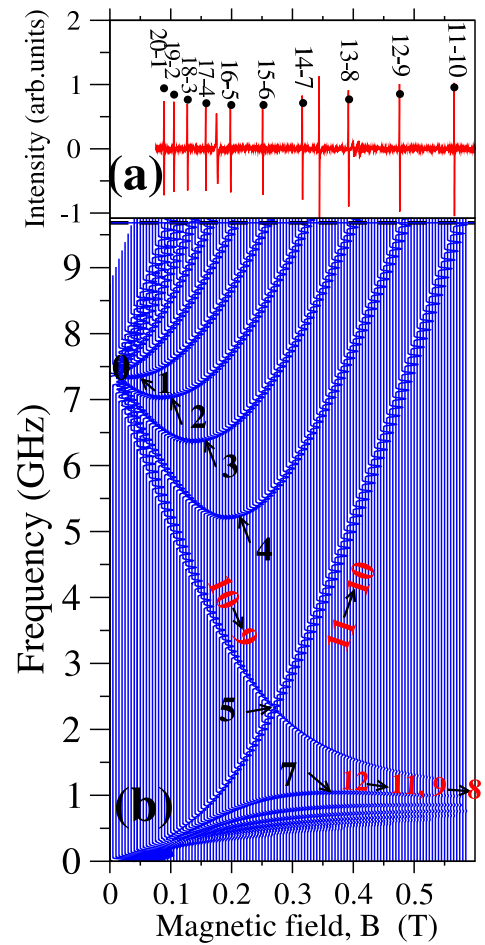} 
\caption{{\bf(a)} Comparison between theory (Eq.\ref{spectra} and \ref{Idip})
(black dots) and experimental CW EPR signal (red online) 
 at 9.7 GHz. Resonances without black dots above them 
are not due to Si:Bi; the large sharp resonance at $0.35$ T is due to
silicon dangling bonds while the remainder are due to defects in the
sapphire ring used as a dielectric microwave resonator. 
The variation in relative intensities is mainly due to the mixing of states as 
in Eq.\ref{basis}. The variability is
not too high but the  calculated intensities are consistent with
experiment and there is excellent agreement for the line positions. 
 {\bf (b)} Calculated EPR spectra
(convolved with the 0.42mT measured linewidth); they are seen to line-up with the
experimental spectra at $f=\frac{\omega}{2\pi}=9.7$ GHz).
The `` cancellation resonances'' are indicated by arrows and integers $-m=0,1,2,3,4...$.
 The $\omega_0=7A$ maximum of $\frac{df}{dB}$
 indicated by the integer 7  corresponds to
that shown in the $\lesssim 2$ GHz ENDOR spectra of \cite{Morton}.} 
\label{Fig2} 
\end{figure}
 It is standard practice  to represent two-state quantum systems using vectors 
 on the Bloch sphere \cite{Schweiger}. We define a parameter 
$R_m^2 = \left[m+{\tilde \omega_0}(1+ \delta)\right]^2 +25-m^2$ where
$R_m$ represents the vector sum magnitude of  spin $x$ and $z$ components.
Denoting $\theta$ as the inclination to the $z$-axis,
  $\cos \theta_m=\left[m+{\tilde \omega_0}(1+ \delta)\right]/R_m$ and $\sin \theta_m=(25-m^2)^{1/2} /R_m$;
then Eq.\ref{Hm}  can also be written:   
\begin{equation} 
{\tilde h}_m= R_m\cos \theta_m {\hat \sigma}_z +R_m\sin \theta_m{\hat \sigma}_x-\frac{1}{2}(1+4{\tilde \omega_0}m \delta){\bf I}. 
\label{Hm1} 
\end{equation} 
  Straightforward diagonalisation
gives the pair of eigenstates, for each $m$, at arbitrary magnetic fields $\omega_0$: 
\begin{align} 
 |\pm, m\rangle &= a^{\pm}_m |\pm\frac{1}{2}, m\mp\frac{1}{2} \rangle +  b^{\pm}_m |\mp\frac{1}{2}, m\pm\frac{1}{2}\rangle
\label{basis} 
\end{align} 
where:
\begin{equation} 
a^\pm_m=\frac{1}{\sqrt{2}}(1+\cos \theta_m)^{1/2}; \ \ 
 b^\pm_m=\pm \frac{1}{\sqrt{2}}\frac{\sin \theta_m}{(1+\cos \theta_m)^{1/2}} 
\label{coeff} 
\end{equation}
and the corresponding eigenenergies:
\begin{equation} 
E^{\pm}_m (\omega_0) = \frac{A}{2}\left[-\frac{1}{2}(1+4{\tilde \omega_0}m \delta) \pm R_m \right]. 
\label{spectra} 
\end{equation} 
 In Fig.\ref{Fig1} the simple (but exact) expression Eq.\ref{spectra} reproduces the 
spin spectra investigated in \eg \cite{LCN} and \cite{Thewalt}. 
Eqs.\ref{coeff} are  
valid for all states except the unmixed $m=\pm 5$ states ($|10\rangle$ and $|20\rangle$).
 For $|m|=5$,  there is no ${\hat \sigma}_x$ 
coupling: these two states are unmixed for {\em all magnetic fields}, thus 
$a_{\pm 5}=1$ and $b_{\pm 5}=0$ and Eq.\ref{spectra}  simplifies drastically to give
$E_{m=\pm 5} =\pm\frac{1}{2}\omega_0 \mp \frac{9}{2}\omega_0 \delta + \frac{9A}{4}$.

 For the doublets, the $a^\pm_m$ are the dominant coefficients at high-field. Then, the
 $B \to \infty$ limit corresponds 
to angle $\theta_m=0$, so $a^\pm_m \to 1$ and $b^\pm_m \to 0$ and the 
states become uncoupled. The $|m| \leq 4$ cancellation resonances  correspond 
to $\theta_m =\pi/2$ so $a^\pm_m=\frac{1}{\sqrt{2}}$, while $b^\pm_m=\pm\frac{1}{\sqrt{2}}$.
 The EPR emission transitions   
$|+,m\rangle \to |-,m-1\rangle$ are dipole allowed at all 
fields. Their 
relative intensities\\ $I \propto 2 I^{+ \to -}_{m \to m-1} |\langle m_s=\frac{1}{2}|S_x|m_s'=-\frac{1}{2}\rangle|^2$.
Since $|\langle \frac{1}{2}|S_x|-\frac{1}{2}\rangle|^2=\frac{1}{2}$,
 variations in line intensities arise from mixing of the states. Thus,
 \begin{equation} 
I^{+ \to -}_{m \to m-1} \propto |a^+_m|^2 |a^-_{m-1}|^2 = \frac{1}{4} (1+ \cos \theta_m)(1+\cos \theta_{m-1}). 
\label{Idip} 
\end{equation}

If mixing is significant $|+,m\rangle \to |+,m-1\rangle$ transitions (of intensity $I^+_m$)
and  $|-,m-1\rangle \to |-,m\rangle$ transitions (of intensity $I^-_m$), EPR-forbidden at high field, become strong,   
 with relative intensities:  
\begin{equation}
 I^{\pm}_{m}=  \frac{1}{4} (1\pm \cos \theta_m)(1 \mp \cos \theta_{m-1}). 
\label{Iforb} 
\end{equation} 
Forbidden lines disappear at high fields;
as $\omega_0 \to \infty$, one can see from Eq.\ref{coeff} that
  $I_{\pm, m\to m-1} \sim \frac{1}{\omega_0^2}   \to 0$ since  $|b^-_{m}|^2 \propto \frac{1}{\omega_0^2}$
at high fields.  
 Eqs.\ref{spectra},  \ref{Idip} and \ref{Iforb}  are exact so are in  
complete agreement with numerical diagonalisation of the full Hamiltonian. 

In Fig.\ref{Fig2}a we test the equations against experimental spectra 
at  $\lesssim 0.6$ Tesla and microwave frequency 9.67849 GHz. The long spin relaxation times at
 low temperatures means that the EPR spectra are easily saturated, complicating the analysis
 of the line intensities. We therefore used a temperature of 42 K so as to measure unsaturated
 resonances. The shorter relaxation times at these elevated temperatures may be due to the presence 
of significant numbers of conduction electrons that are no longer bound to Si:Bi donors. We measure
 a very broad microwave absorption centred on zero magnetic field (subtracted from Fig.\ref{Fig2}a)
which we attribute to these 
conduction electrons. 
 The comparison with experiment shows excellent agreement with the positions of the resonances,
which are far from equally spaced in the low-field regime. 
 Experimental lines are found to be Gaussians of width $\approx 0.42$mT; this was 
 attributed in \cite{LCN} to the effects of  
$^{29}$Si in this sample; samples with enriched $^{28}$Si are expected to give much narrower 
linewidths.  
 
For Fig.\ref{Fig2}b, we generated the spectra, convolved with $0.42$mT Gaussians to obtain 
the EPR spectra of all lines (both allowed and forbidden) at all frequencies below 10GHz. 
We indicate the main dipole allowed lines as well as indicating the approximate position 
of the main resonances. The spectra show a striking landscape of transitions which 
show maxima or minima where $\frac{df}{dB}=0$ and double-valued EPR resonant fields 
(\ie transitions with EPR resonances at two different magnetic fields).
 No Boltzmann factor has been included in the simulation.
  The nuclear field splittings are unresolved and extremely 
small; they do not affect line intensities significantly. 
 To simplify our discussion, we neglect the tiny nuclear shifts $\propto \omega_0 \delta$, but include them 
whenever spectroscopically significant.

The well-studied 4-state $S=1/2$, $I=1/2$ Si:P system can be mapped
onto a two-qubit basis.
With a 20-eigenstate state-space, the Si:Bi spectrum is more
complex, but we can identify a natural subset of 4 states, which
represents an effective 2-coupled-qubit analogue:
\begin{eqnarray}
 &|12\rangle& \to (1)_e(1)_n \nonumber \\
 &|11\rangle& \to (1)_e(0)_n \nonumber \\
 &|9\rangle&  \to (0)_e(1)_n \nonumber \\
&|10\rangle& \to (0)_e(0)_n \nonumber 
\label{cbasis}
\end{eqnarray}
 As hyperpolarization initialises the spins in state 
$|10\rangle $ and this state has both the electron and nuclear spins
fully anti-aligned with the magnetic field,
 it can be identified with the $(0)_e(0)_n$ state. The other states are related to
it by either one or two qubit flips, just as in the Si:P basis.

To have a universal set of gates for quantum information it is known to be sufficient
 to be able to perform arbitrary single qubit manipulations and
 a control-NOT (CNOT) gate \cite{DiVincenzo}. In the two qubit system described here,
 arbitrary electronic qubit-only manipulations can be performed with radiation
 pulses exciting transitions between states $12\to 9$ and $10\to 11$,
while single nuclear-qubit rotations correspond to $12\to 11$ and $10\to 9$.
  The CNOT gate (for example using the nuclear spin as a control qubit) is even simpler as it requires
only a $\pi$ pulse connecting $12\to 9$ \cite{long}. 

The electronic flips are EPR allowed at all fields for
both Bi and P donors so can be performed in a time on the order of 10 ns \cite{LCN}.
 The nuclear transitions, however, 
 require a slower,
(of order microseconds) NMR pulse for Si:P. For Si:Bi, on the other hand, at the
$m=-4$ resonance the nuclear and electronic transition strengths become exactly equal
as may be verified by setting $\theta_{-5}=0$ and $\theta_{-4}=\pi/2$ in 
Eqs.(\ref{Idip}) and (\ref{Iforb}). Time-dependent calculations \cite{long} show that
the duration of a $\pi$ pulse is also equalized.

 \begin{figure}[] 
\includegraphics[width=3.5in]{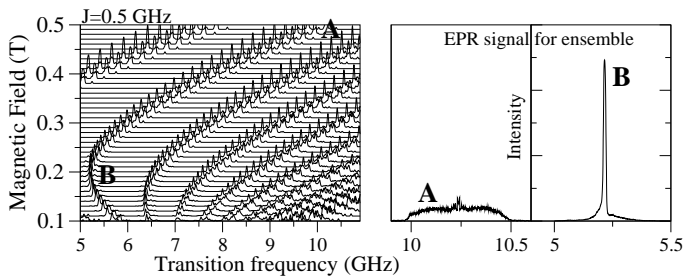} 
\caption{The simulated EPR spectra for pairs of interacting Si:Bi donors shows ultra-narrow 
lines at cancellation resonances,
somewhat analogous to ``exact cancellation'' in ESEEM spectroscopy.
The electronic spins 
are coupled by an exchange term $J {\bf {\hat S}_1 . {\hat S}_2}$. The left panel shows the effect of 
a typical $J=0.5$ GHz coupling. A splitting (of order $J$) appears in general, but 
 near the $m=-4,-3$ resonances, this is suppressed. The graphs on the right show the
calculated signal from a sample of spins with a distribution of $J$,
where average $\langle J\rangle=0.3$ GHz, variance $\sigma_J=0.3$.
At the cancellation resonances, despite the ensemble averaging, the lines remain narrow.}   
\label{Fig3} 
\end{figure} 

 {\em The $\omega_0=A,2A,3A,4A$ resonances} yield textbook level anti-crossings
 as well as ``Bell-like'' eigenstates.
 Were it to become possible to vary the external field 
sufficiently fast to produce sudden, rather than adiabatic evolution of the eigenstates
 it would be possible to transfer the former to the high-field regime.
Unfortunately, ramping magnetic fields (up or down) sufficiently fast to violate adiabaticity,
 though not impossible, would require some of the fastest magnetic field pulses ever 
 produced (eg $10^8$T/s obtained by \cite{Boebinger}). 
 However, we show that adiabatic magnetic field sweeps already achievable by
 ordinary laboratory magnetic pulses ($1-10$ T/ms) suffice to already achieve new
possibilities.\\

 {\em The frequency minima at 5-8 GHz: } fields at which  $df/dB=0$ 
are expected to lead to a reductions of decoherence,
since sensitivity to magnetic fluctuations is minimised;
a measurable reduction has been seen \cite{Morton} by varying the ratio $df/dB$ by 
$30-50\%$. In Fig.(\ref{Fig2}) we see several transitions have a minimum frequency.
 These minima  (in effect of $R_m+R_{m-1}$) occur for: 
\begin{equation} 
 \cos \theta_m=  -\cos \theta_{m-1}. 
\label{min} 
\end{equation}
Thus  $\theta_m= \pi-\theta_{m-1}$; the consequence is that 
the minima lie exactly midway {\em in angular coordinates} between cancellation resonances. 
For example, for the $12\to 9$ line ($|+,m=-3\rangle \to |-,m=-4\rangle$) 
the minimum is at
$\omega_0= \frac{25A}{7}= 3.57A$ so $B\simeq 0.188$ T.
 Here, the $m=-4$ doublet has passed its resonance point at $0.21$ T
(for which $\theta_{-4}= \pi/2$) by an angle $\phi= \arccos \frac{21}{15\sqrt{2}}$
and the $m=-3$ resonance is at an equal angular distance {\em before} its resonance at $\simeq 0.16$ T:
thus $\theta_{-4}= \pi/2+ \phi$ while $\theta_{-3}= \pi/2- \phi$.
Both doublets are quite close to the Bell-like form.

{\em Line narrowing: } an interesting and unexpected consequence has applications to
studies with larger numbers of spins. A pair 
of Bi atoms, interacting via a spin-exchange term of the form
$J {\bf {\hat S}_1 . {\hat S}_2}$ will result in splitting of the EPR spectral lines (with an
energy splitting of order $J$). However this is suppressed near the cancellation resonances.
Analogously to ``exact cancellation'', this
 makes the system less sensitive to ensemble averaging.
 For exact cancellation, this means the averaging over
different orientations in powder spectra; here it means magnetic perturbations including
spin-spin interactions.   Fig.\ref{Fig3} (left panel) plots the signal for $J=0.5$ GHz for a single
pair of Si:Bi atoms and clearly shows the line splitting away from the 
resonances. The right panels show the effects of averaging many spectra
each corresponding to different $J$ (with an average $\langle J \rangle =0.3$ GHz
and width $\sigma_J=0.3$ GHz). While typical spectra show 
a broad feature of width $\sim \sigma_J$, at the cancellation resonance the line width remains
strikingly narrow (close to the single atom line width).

{\em A frequency maximum } at ${\tilde \omega_0}= 7$ and $B\simeq 0.37$ T is
 marked with a $7$ in Fig.\ref{Fig2}b. 
We can show that a $df/dB=0$ point which is a  {\em maximum} (in effect of $R_m-R_{m-1}$) implies: 
\begin{equation} 
 \cos \theta_m= \cos \theta_{m-1}. 
\label{max} 
\end{equation}
This condition  does not correspond to the elimination of 
 the field splitting terms; instead it implies
 $\theta_{-3}= \theta_{-4}=\pi/4$, thus equalizing the Bloch angle for the
associated energy levels. In this sense it is somewhat different
to the other cancellation resonances; nevertheless, it still provides 
a $df/dB=0$ point and thus some potential for reducing broadening and decoherence.
In \cite{long}, it is shown that the ${\tilde \omega_0}= 7$ resonance offers new possibilities for
copying and storing qubit states.
 At ${\tilde \omega_0}= 5$, the most drastic cancellation resonance
occurs, since both $\sigma_z$ and $\sigma_x$ terms in Eq.\ref{Hm} are eliminated,
leaving only the isotropic term. Although there is no $df/dB=0$ or line narrowing here,
there is a possibility of driving, by a second order process, simultaneous qubit
rotations \eg $ |0\rangle_e|1\rangle_n \to |1\rangle_e|0\rangle_n$  see \cite{long}.

{\em Conclusions}: In the intermediate-field regime ($B \simeq 0.05-0.6$ T)
the exceptionally large values of $A$ and $I$ for Si:Bi
generate a series of  cancellation resonances.
 They are associated not only with level crossing structures but also with more subtle and
not previously studied effects: 
both line broadening and decoherence effects may be reduced; also,
if the electronic and nuclear spins of Si:Bi are used as a 
2-coupled qubit system, the cancellation resonances allow a universal
 set of quantum gates to be performed with fast EPR microwave pulses, eliminating the
need for slower radio frequency addressing of the nuclear qubit.
One scheme would envisage the following stages:
(1) hyperpolarization of the sample into state $|10\rangle$ (in which
the 2-qubit system is initialized as $|0\rangle_e|0\rangle_n$) at
 $B \approx 5 $T. (2) A magnetic field pulse ($\sim 10$ T/ms, of duration lower than decoherence times)
 would reduce $B$ to $\simeq 0.1$T. (3) As the pulse ramps up, a series of EPR pulses would execute a series
  of gates and operations on the system.
 (4) As the magnetic pulse decays, the system is restored
 to the high-$B$ limit, leaving it in the desired superposition of 
 $|0\rangle_e|0\rangle_n,|1\rangle_e|1\rangle_n, |0\rangle_e|1\rangle_n$ and $|1\rangle_e|0\rangle_n$
 basis states.
 Thus, given the capability to rapidly ($\lesssim 1$ms) switch from the high to intermediate
field regime, Si:Bi confers significant additional possibilities for quantum information
 processing relative to Si:P.


\begin{thebibliography}{99}
\bibitem{ESEEM} ENDOR is electron-nuclear double resonance; 
 ESEEM is electron spin echo envelope eodulation.
\end{thebibliography}
\end{document}